\documentclass[12pt, preprint]{aastex}

\usepackage{amssymb}
\usepackage{amsmath}
\usepackage{epsfig}
\usepackage{graphicx}
\usepackage{natbib}
\usepackage{multirow}
\usepackage{rotating}
\usepackage{longtable}
\usepackage{verbatim}
\usepackage{url}
\usepackage{hyperref}
\usepackage{caption2}
\usepackage{CJK}

\newcommand{\mearth}{$\rm M_{\earth}$}

\begin{document}
\begin{CJK*}{UTF8}{gbsn}

\title{
A Simple Analytical Model for Rocky Planet Interiors
}
\author{Li Zeng\altaffilmark{1,a} and Stein B. Jacobsen\altaffilmark{1,b}
}
\affil{$^1$Department of Earth and Planetary Sciences, Harvard University, Cambridge, MA 02138}

\email{$^a$astrozeng@gmail.com} 
\email{$^b$jacobsen@neodymium.harvard.edu}

\begin{abstract}

This work aims at exploring the scaling relations among rocky exoplanets. With the assumption of internal gravity increasing linearly in the core, and staying constant in the mantle, and tested against numerical simulations, a simple model is constructed, applicable to rocky exoplanets of core mass fraction (CMF) $\in 0.1\sim0.4$ and mass $\in 0.1 \sim 10 M_{\oplus}$. Various scaling relations are derived: (1) core radius fraction $\rm CRF \approx \sqrt{CMF}$, (2) Typical interior pressure $P_{\text{typical}} \sim g_{\rm{s}}^2$ (surface gravity squared), (3) core formation energy $E_{\rm diff} \sim \frac{1}{10} E_{\rm grav}$ (the total gravitational energy), (4) effective heat capacity of the mantle $C_{\rm p}\approx \left( \frac{M_{\rm{p}}}{M_{\oplus}} \right) \cdot 7 \cdot 10^{27}$ J K$^{-1}$, and (5) the moment of inertia $I\approx \frac{1}{3} \cdot M_{\rm{p}} \cdot R_{\rm{p}}^2$. These scaling relations, though approximate, are handy for quick use owing to their simplicity and lucidity, and provide insights into the interior structures of rocky exoplanets. As examples, this model is applied to several planets including Earth, GJ 1132b, Kepler-93b, and Kepler-20b, and made comparison with the numerical method.

\end{abstract}

\section{Introduction}

\begin{figure}
\centering
\includegraphics[width=4.5in]{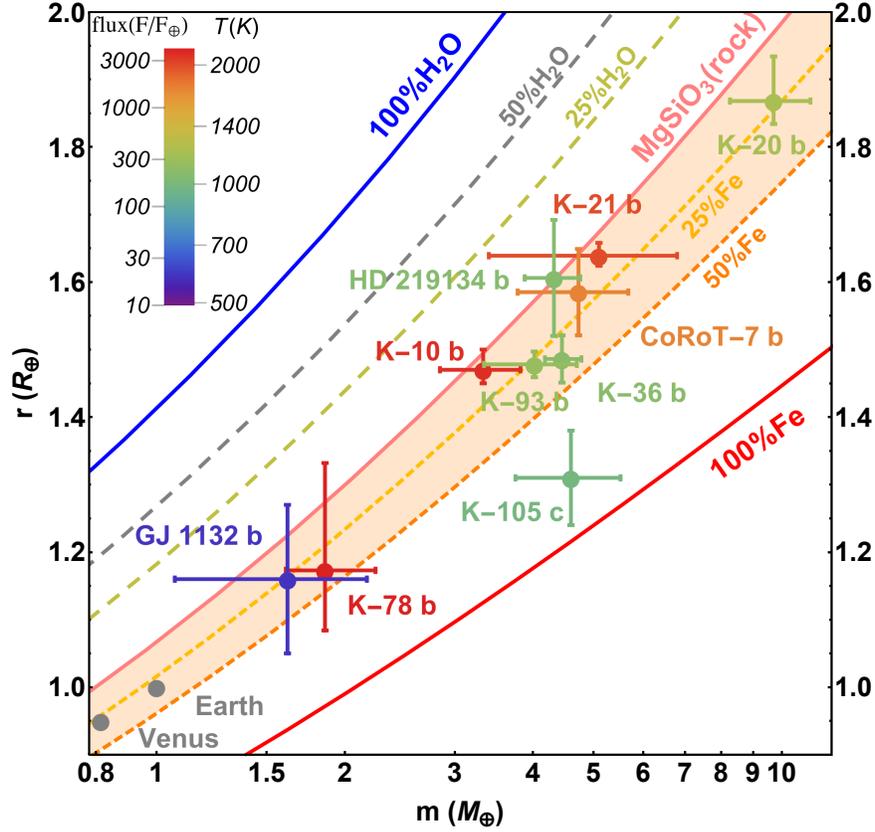}
\caption{\singlespace Mass-radius plot showing selected rocky planets. Curves show models of different compositions, with solid indicating single composition (Fe, $\rm MgSiO_3$, i.e. rock, $\rm H_2O$) and dashed indicating Mg-silicate planets with different amounts of $\rm H_2O$ or Fe added. Rocky planets without volatile envelope likely lie in the shaded region within uncertainty, and those ones with volatile envelope may lie above. Planets are color-coded by their incident bolometric stellar flux (compared to the Earth) and equilibrium temperatures assuming (1) circular orbit (2) uniform surface temperature (3) bond albedo A=0. Earth and Venus are shown for reference. }
\label{figure_mrplot}
\end{figure}

Masses and radii of rocky exoplanets have been found for about a dozen cases (Figure~\ref{figure_mrplot}), including Kepler-21b~\citep{Howell:2012,Lopez-Morales:2016}, Kepler-20b~\citep{Gautier:2012,Buchhave:2016}, COROT-7b~\citep{Leger:2009,Queloz:2009,Hatzes:2011,Wagner:2012,Haywood:2014,Barros:2014}, HD219134b~\citep{Vogt:2015,Motalebi:2015}, Kepler-10b~\citep{Batalha:2011,Wagner:2012,Dumusque:2014,Esteves:2015}, Kepler-93b~\citep{Ballard:2014,Dressing:2015}, Kepler-36b~\citep{Carter:2012,Morton:2016}, Kepler-78b~\citep{Pepe:2013,Grunblatt:2015}, K-105c~\citep{Rowe:2014,Holczer:2016,Jontof-Hutter:2016}, GJ 1132b~\citep{Berta-Thompson:2015,Schaefer:2016}, and more are likely. Here we explore what other parameters can be further gleaned from this information. Our earlier work~\citep{Zeng:2016} shows that by using an equation of state (EOS) for Earth for different core mass fractions (CMFs), a simple relationship between CMF, planetary radius, and mass can be found as

\begin{equation}
\text{CMF} = \frac{1}{0.21} \cdot \left[1.07-\left(\frac{R_{\rm p}}{R_\oplus}\right)/\left(\frac{M_{\rm p}}{M_\oplus}\right)^{0.27}\right]
\label{eq:cmf0}
\end{equation}

This work shows that the CMF can be related to the core radius fraction (CRF) of a rocky planet. A simple structural model can be calculated, which depends only on three parameters, (1) surface gravity $g_{\rm{s}}$, (2) planet radius $R_{\rm{p}}$, and (3) core radius fraction CRF. The procedure is as follows: 

\begin{enumerate}
\item Surface gravity $g_{\rm{s}}=\frac{G \cdot M_{\rm{p}}}{R_{\rm{p}}^2}$ can be calculated from mass $M_{\rm{p}}$ and radius $R_{\rm{p}}$ of a rocky planet, or directly from combining transiting depth with radial-velocity amplitude (Equation~(\ref{eq: gs})).
\item CMF can be determined from Equation~(\ref{eq:cmf0}). 
\item CRF can be estimated as $\sqrt{\text{CMF}}$. 
\end{enumerate}

\begin{figure}
\centering
\includegraphics[width=7.0in]{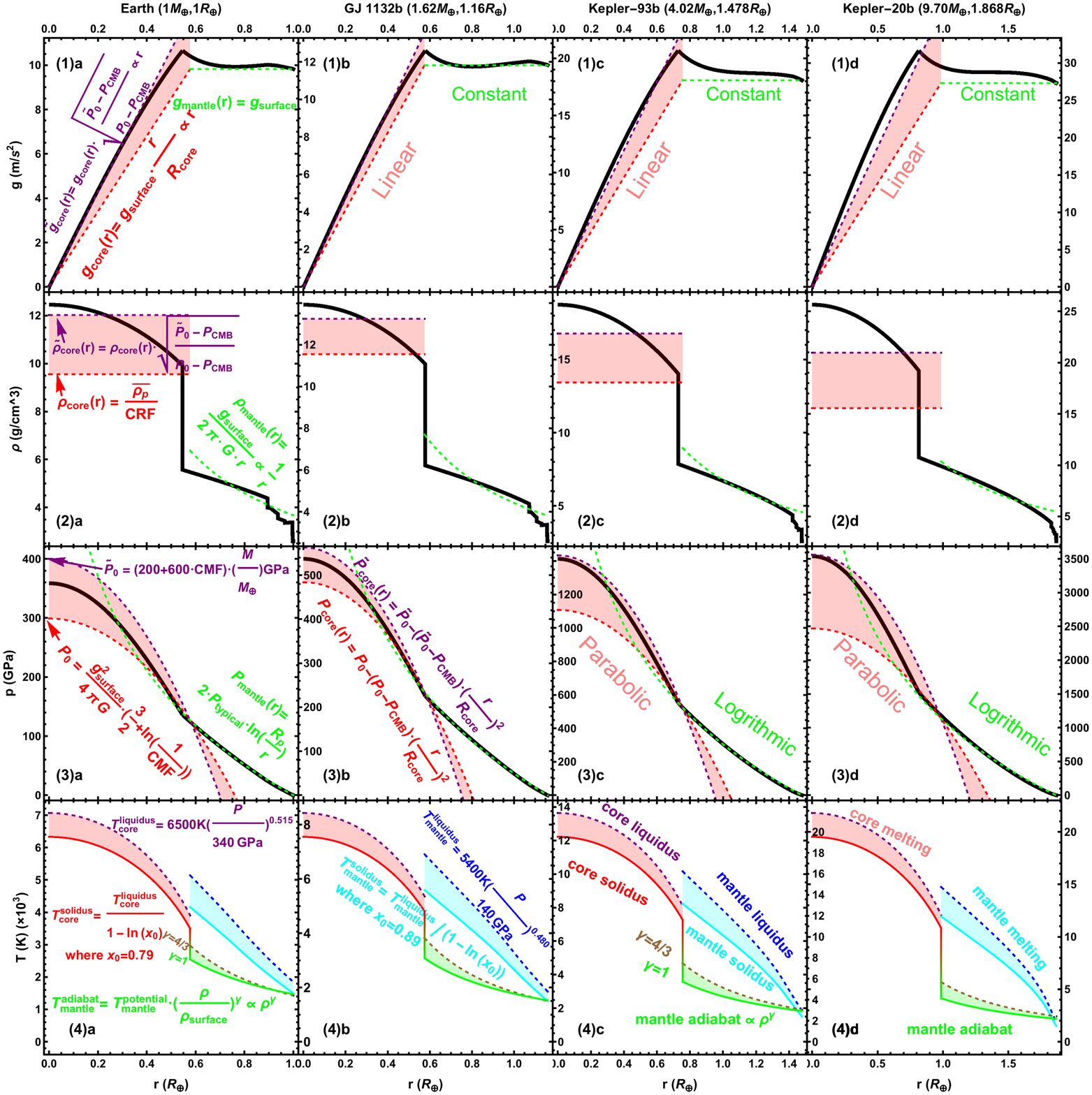}
\caption{\singlespace Numerical calculations based on PREM-extrapolated EOS (black) versus simple analytical models: Core (red, purple, and pink-area in between) and Mantle (green). Panel (1-4)a: Earth, Panel (1-4)b: GJ 1132b, Panel (1-4)c: Kepler-93b, Panel (1-4)d: Kepler-20b. Panel (1)a-d: Gravity Profiles (core is proportional to $r$ and mantle is constant). Panel (2)a-d: Density Profiles (core is constant and mantle is inversely proportional to $r$). Panel (3)a-d: Pressure Profiles (core is parabolic in $r$ and mantle is logarithmic in $r$). Panel (4)a-d: Temperature Profiles (best estimates shall lie in the green area (mantle) and pink area (core)). The solidus (where mixture starts to melt) and liquidus (where mixture completely melts) are plotted for comparison. }
\label{figure_profiles}
\end{figure}

The only assumption of this model is that the internal gravity profile can be approximated as a piecewise function (see Figure~\ref{figure_profiles} Panel (1)a-d):
\begin{enumerate}
\item In the core, the gravity $g$ increases linearly with radius from 0 at the center to $g_{\rm{s}}$ (surface value) at the core-mantle boundary (CMB): $g_{\text{core}}(r) = g_{\rm{s}} \cdot \left( \frac{r}{R_{\text{core}}} \right) \propto r$
\item In the mantle, $g$ stays constant: $g_ {\text{mantle}}(r) = g_{\rm{s}}=\text{const}$
\end{enumerate}

\noindent This assumption is equivalent to assuming constant core density, followed by the mantle density decreasing to $\frac{2}{3}$ of the core density at CMB, and density deceasing as $\frac{1}{r}$ in the mantle.

The validity of this assumption is tested against the numerical results from solving the planetary structures with realistic EOS derived from PREM~\citep{Dziewonski:1981}, across the mass-radius range of 0.1$\sim$10~\mearth~and 0.1$\lesssim$CMF$\lesssim$0.4 for two-layer (core+mantle) rocky planets. Note that Mercury lies outside this range of the CMF as it has a big core, owing to its likely giant impact origin~\citep{Asphaug:2014}.

Various scaling relations are derived from this model.

\section{Scaling Relation between Pressure and Gravity}

The two first-order differential equations~\citep{Seager:2007,  Zeng_Seager:2008, Zeng_Sasselov:2013} governing rocky planet interiors are

1. a hydrostatic equilibrium (force balance) equation: 
\begin{equation}
\frac{dP}{dr} = -\frac{Gm\rho}{r^2} =-g\cdot \rho
\label{eq:1}
\end{equation}

2. a mass conservation equation: 
\begin{equation}
\frac{dm}{dr} = 4 \pi r^2 \rho
\label{eq:2}
\end{equation}

Equations~(\ref{eq:1}) and~(\ref{eq:2}) can be combined to give a relation between the internal pressure $P$ and the mass $m$ (mass contained within radius $r$, now used as the independent variable instead): 

\begin{equation}
\frac{dP}{dm} = -\frac{G m}{4\pi r^4} = -\frac{1}{4\pi G} \cdot \frac{g^2}{m} = -\frac{1}{4\pi G} \cdot g^2 \cdot \frac{d\ln (m)}{dm}
\label{eq:P2}
\end{equation}

\noindent Integrating Equation~(\ref{eq:P2}), we get ($\ln(m)$ stands for the natural logarithmic of $m$)

\begin{equation}
\boxed{\int_{\text{surface}}^{\text{interior}} dP = -\frac{1}{4\pi G} \int_{M_{\rm{p}}}^{\text{mass enclosed inside}} g^2 \cdot d \ln (m)}
\label{eq:P3}
\end{equation}

This integration is from the surface inward, as the pressure at the surface is zero.
Therefore, the typical internal pressure is on the order of

\begin{equation}
P \sim \frac{\overline{{g}^2}}{4\pi G}
\label{eq: Pscaleg2}
\end{equation}

\noindent where $\overline{{g}^2}$ is some average of $g^2$. Defining planet mass as $M_{\rm{p}}$, planet radius as $R_{\rm{p}}$, and planet mean density as $\overline{\rho_p}$, then surface gravity $g_{\rm{s}}$ and characteristic interior pressure $P_{\text{typical}}$ are

\begin{equation}
g_{\rm{s}} \equiv \frac{GM_{\rm{p}}}{R_{\rm{p}}^2}
\label{eq:gsurface}
\end{equation}

\begin{equation}
P_{\text{typical}} \equiv \frac{{g_{\rm{s}}}^2}{4\pi G} = \frac{GM_{\rm{p}}^2}{4\pi R_{\rm{p}}^4}
\label{eq:Ptypical}
\end{equation}

\noindent $P_{\text{typical}}$ will be shown to approximate $P_{\text{CMB}}$ (pressure at CMB) later on. Given $g_{\rm{s}}$ in S.I. units ($\rm m~s^{-2}$) and pressures in GPa: 

\begin{equation}
\boxed{ P_{\text{CMB}} \sim P_{\text{typical}} \sim g_{\rm{s}}^2 }
\label{eq:P}
\end{equation}

\noindent For example, $g_{\oplus}$ (Earth's gravity)$\approx10\rm m~s^{-2}$, and $g_{\oplus}^2 \approx 100$ is near $P_{\oplus,\text{CMB}}=136$ GPa. The values of other planets are listed in Table~\ref{Table1}. 

\section{Density Profile}

Based on the assumption of the gravity profile, the density profile is

\begin{subequations}
\begin{equation}\label{eq:densitycore}
\rho_{\text{core}}(r) =  \frac{3 g_{\rm{s}}}{4 \pi G R_{\text{core}}} =\frac{ \overline{\rho_p} }{\text{CRF}} = \text{constant, thus, } m_\text{core}(r) \propto r^3
\end{equation}

\begin{equation}\label{eq:densitymantle}
\rho_{\text{mantle}}(r) = \frac{g_{\rm{s}}}{2 \pi G  r} \propto \frac{1}{r} \text{, thus, } m_\text{mantle}(r) \propto r^2
\end{equation}
\end{subequations}

Figure~\ref{figure_profiles} Panel (2)a-d compares this to the PREM-derived density profiles. The $\frac{1}{r}$ dependence approximates the compression of mantle material toward depth, and the smaller core (CMF $\lesssim$0.4) allows the core density to be approximated as constant. Generally, Equation~(\ref{eq:densitycore}) approximates the density of the core near the CMB. Anywhere in the mantle, 

\begin{equation}
\frac{m}{M_{\rm{p}}}=\left( \frac{r}{R_{\rm{p}}} \right)^2
\label{eq:mr3}
\end{equation}

\noindent In particular, at the CMB,  

\begin{equation}
\text{CMF} = \frac{M_{\text{core}}}{M_{\rm{p}}} = \left( \frac{R_{\text{core}}}{R_{\rm{p}}} \right)^2 = \text{CRF}^2
\label{eq:cmf1}
\end{equation}

\noindent In reality, this exact relation (Equation~(\ref{eq:cmf1})) becomes approximate:
\begin{equation}
\boxed{\text{CRF} \approx \sqrt{\text{CMF}}}
\label{eq:cmf}
\end{equation}

The error of Equation~(\ref{eq:cmf}) is generally within $\sim$10\% (see Table~(\ref{Table1})). It is a quick way to estimate the CRF from the CMF and vice versa. It can even be applied to a rocky planet with a volatile envelope if it is only applied to the solid portion of that planet.

\begin{table}[htbp]

		\caption{Calculated parameters for four planets from this simple analytical model}

			\centering
			\scalebox{0.8}{
			
			\begin{tabular}{| c | c | c | c | c |}
			
			\hline

				                           & Earth & GJ 1132b & Kepler-93b & Kepler-20b \\

				\hline

				M(M$_{\oplus}$) & $1$ & $1.62$ & $4.02$ & $9.70$ \\
			
				R(R$_{\oplus}$) & $1$ & $1.16$ & $1.478$ &  $1.868$ \\
				
	                      	CMF (Equation~(\ref{eq:cmf0})) & $0.33$& $0.25$ & $0.26$ & $0.28$ \\
		
				CMF$_{\rm N}$\footnotemark[1] & $0.32$ & $0.27$ & $0.28$ & $0.22$  \\
	
				CRF (Equation~(\ref{eq:cmf})) & $0.58$ & $0.50$ & $0.51$ & $0.53$  \\
		
				CRF$_{\rm N}$ & $0.55$ & $0.49$ & $0.49$ & $0.44$  \\

				$\overline{\rho_p} (\rm g~cm^{-3})$ & $5.5$ & $5.7$ & $6.9$ & $8.2$ \\
				
				
				

			        $g_{\rm s} (\rm m~s^{-2})$ (Equation~(\ref{eq:gsurface})) & $9.8$ & $12$ & $18$ & $27$ \\
	 
				$P_{\rm typical}$ (TPa\footnotemark[2]) (Equation~(\ref{eq:Ptypical})) & $0.12$ & $0.17$ & $0.4$ & $0.9$ \\
				
				$P_{\rm CMB}$ (TPa) (Equation~(\ref{eq:Pcmb})) & $0.13$ & $0.23$ & $0.5$ & $1.1$ \\
				
				$P_{\rm 0}$ (TPa) (Equation~(\ref{eq:P0_1}) & $0.3$ & $0.5$ & $1.1$ & $2.5$ \\

				$\widetilde{P_{\rm 0}}$ (TPa) (Equation~(\ref{eq:P0_tilda})) & $0.40$ & $0.56$ & $1.4$ & $3.6$ \\

				$E_{\rm grav} (10^{32} \rm J)$ (Equation~(\ref{eq:Egrav})) & $2.4$ & $5.6$& $27$ & $120$ \\
												
				$E_{\rm diff} (10^{32} \rm J)$ (Equation~(\ref{eq:Ediff1})(\ref{eq:Ediff2})) & $0.2$ & $0.5$ & $2.4$ & $11$ \\
								
				$E^{\rm thermal}_{\rm mantle} (10^{32} \rm J)$ (Equation~(\ref{eq:mantleEth})(\ref{eq:mantleEth2})) & $0.1$ & $0.2$ & $0.5$ & $1.1$ \\
				
				$T_{\rm mp} (10^{3} \rm K)$ & $1.6$ & $1.6$ & $1.6$ & $1.6$ \\
				
				$T^{\rm CMB}_{\rm mantle} (10^{3} \rm K)$ & $2.5$ & $3.1$ & $3.6$ & $4.1$ \\
				
				$T^{\rm CMB}_{\rm core} (10^{3} \rm K)$ & $4$ & $5$ & $8$  & $11$ \\
				
				$T_{\rm center} (10^{3} \rm K)$ & $6$ & $7$ & $12$ & $20$ \\

			         $I (10^{38} \rm kg~m^2)$ (Equation~(\ref{eq:MoIcombined})(\ref{eq:moment_of_inertia})) & $0.8$ & $1.8$ & $7.2$ & $28$ \\
				
				\hline 
			
		  \end{tabular}
	}	
	 \label{Table1}	
\end{table}
\footnotetext[1]{N stands for numerical simulation using {\bf ManipulatePlanet} at \url{astrozeng.com}}
\footnotetext[2]{TPa = 1000 GPa = $10^{12}$ GPa. }


\section{Pressure Profile}

\subsection{Pressure in the Mantle}

Integrating Equation~(\ref{eq:P3}) with constant mantle gravity, we obtain

\begin{equation}
P_{\text{mantle}}(m) = \frac{g_{\rm{s}}^2}{4\pi G} \cdot \ln \left( \frac{M_{\rm{p}}}{m} \right) = P_{\text{typical}} \cdot \ln \left( \frac{M_{\rm{p}}}{m} \right) = 2 P_{\text{typical}} \cdot \ln \left( \frac{R_{\rm{p}}}{r} \right)
\label{eq:Pmantle}
\end{equation}

\noindent Evaluating Equation~(\ref{eq:Pmantle}) at the CMB gives $P_{\text{CMB}}$ (pressure at the CMB): 

\begin{equation}
\boxed{P_{\textrm{CMB}} = P_{\text{typical}} \cdot \ln \left( \frac{1}{\text{CMF}}  \right)  = \frac{g_{\rm{s}}^2}{4\pi G} \cdot \ln \left( \frac{1}{\text{CMF}}  \right)}
\label{eq:Pcmb}
\end{equation}

\noindent For CMF $\in 0.1 \sim 0.4$, $P_{\text{CMB}} \in (0.9\sim2.3)P_{\text{typical}}$. 

$P_{\text{CMB}}$ is an important physical parameter, as it determines the state of core and mantle materials in contact. Equation~(\ref{eq:Pcmb}) only depends on $g_{\rm{s}}$ and CMF. And $g_{\rm{s}}$ can be determined independent of the stellar parameters~\citep{Southworth:2007} as

\begin{equation}
g_{\rm{s}} = \frac{2\pi}{P_{\text{orb}}} \frac{(1-e^2)^{1/2}A_{\rm RV}}{(R/a)^2 Sin[i]} 
\label{eq: gs}
\end{equation}

\noindent where semi-amplitude $A_{\rm RV}$ and orbital eccentricity $e$ can be constrained from the radial-velocity curve, and $R/a$ is the radius over the semi-major axis ratio, which could be constrained directly from the transit light curve. The orbital period $P_{\text{orb}}$ can be constrained from both. Thus, it is possible to estimate $P_{\text{CMB}}$ even without knowing the accurate mass and radius in some cases for rocky planets.

\subsection{Range of Applicability of This Model}

From a theoretical point, we explore the range of applicability of this model.

Bulk modulus $K\equiv \frac{\partial P}{\partial \ln(\rho)}$. Therefore, in the mantle,   

\begin{equation}
K_{\text{mantle}} = \frac{\partial P_{\text{mantle}}}{\partial \ln(\rho_{\text{mantle}})} = \frac{g_{\rm{s}}^2}{4\pi G} \cdot \frac{d\ln(m)}{d\ln(r)} = 2 \cdot P_{\text{typical}}
\label{eq:Kmantle}
\end{equation}

Thus, in this model, the bulk modulus is constant everywhere in the mantle, equal to twice the typical internal pressure $P_{\text{typical}}$. Realistically, $K$ shall increase with pressure, so how good is this approximation?

For Earth, $P_{\oplus, \text{typical}} = \frac{{g_{\oplus}}^2}{4\pi G} = 115~\rm GPa$, so $K_{\oplus,\text{mantle}} = 230~\rm GPa$.  Comparing it to the isentropic bulk modulus $K_s$ of Earth's mantle according to PREM: $\begin{cases} 
K_{\oplus, LID} = 130~\textrm{GPa} \\ 
K_{\oplus, 670km} = 255.6\sim 300~\textrm{GPa} \\ 
K_{\oplus, D''} = 640~\textrm{GPa}
\end{cases}$

\noindent So $K_{\oplus,\text{mantle}}$ represents the midrange of the realistic bulk modulus in the mantle. For higher masses, let us invoke the BM2 (Birch-Murnaghan second-order) EOS~\citep{Birch:1947, Birch:1952}, which when fitted to PREM gives $K_0 \approx 200$GPa for both core and mantle~\citep{Zeng:2016}: 

\begin{equation}
P=\frac{3}{2} \cdot K_0 \left[\left(\frac{\rho}{\rho_0}\right)^{\frac{7}{3}}-\left(\frac{\rho}{\rho_0}\right)^{\frac{5}{3}} \right]
\label{eq:BM2}
\end{equation}

\noindent Again, $K$ is obtained by differentiating Equation~(\ref{eq:BM2}): 

\begin{equation}
K \equiv \frac{\partial P}{\partial ln(\rho/ \rho_0)}=\frac{3}{2} \cdot K_0 \left[\frac{7}{3} \cdot \left(\frac{\rho}{\rho_0}\right)^{\frac{7}{3}}-\frac{5}{3} \cdot \left(\frac{\rho}{\rho_0}\right)^{\frac{5}{3}} \right]
\label{eq:K}
\end{equation}

\noindent Equation~(\ref{eq:K}) suggests: $\begin{cases} \text{when } P\lesssim K_0, K \approx K_0\\
 \text{when } P\gg K_0, K \rightarrow \frac{7}{3}P \approx 2P \end{cases}$. Since $P_{\text{typical}}$ is the typical pressure in the mantle, $K \approx 2P \approx 2 P_{\text{typical}}$. It is the same as Equation~(\ref{eq:Kmantle}). Therefore, this approximation shall exist for higher masses as long as BM2 EOS holds. BM2 EOS's validity range extends above 10 $M_\oplus$, and here we set the upper limit to be 10 $M_\oplus$. 

\subsection{Pressure in the Core}

Since $\rho_{\text{core}}$ = constant in this approximation, from Equation~(\ref{eq:1}) we have 

\begin{equation}
\frac{dP_{\rm core}(r)}{dr} = -\frac{G}{r^2}\cdot \left( \frac{g_{\rm{s}}}{GR_{\rm core}}r^3 \right)\cdot \left( \frac{3g_{\rm{s}}}{4\pi G R_{\rm core}} \right) = -\frac{3r}{R_{\rm core}^2} \cdot P_{\text{typical}}
\end{equation}

Integrating it gives the pressure dependence on the radius as a parabolic function: 

\begin{equation}
P_{\text{core}}(r) = P_0 - \frac{3}{2} \cdot P_{\text{typical}} \cdot \left( \frac{r}{R_{\text{core}}} \right)^2
\label{eq:Pcore}
\end{equation}

$P_0$ (central pressure) can be determined by connecting Equation~(\ref{eq:Pcore}) at CMB to $P_{\text{CMB}}$ from Equation~(\ref{eq:Pcmb}) as

\begin{equation}
P_0 = P_{\text{CMB}} + \frac{3}{2}P_{\text{typical}} = P_{\text{typical}} \cdot \left[ \ln \left( \frac{1}{\text{CMF}} \right) +\frac{3}{2} \right] \Rightarrow P_0 \in (2.4\sim3.8) \cdot P_{\text{typical}} \label{eq:P0_1}
\end{equation}

Therefore, in this approximation, the pressure dependence on the radius is piecewise: parabolic in the core (Equation~(\ref{eq:Pcore})) and logarithmic in the mantle (Equation~(\ref{eq:Pmantle})), and they interconnect at CMB. This piecewise pressure profile can be closely matched to the realistic pressure profile calculated from PREM as shown in Figure~\ref{figure_profiles} Panel (3)a-d. Equation~(\ref{eq:P0_1}) tends to significantly underestimate the $P_0$ towards higher mass and higher CMF due to significant core compression.  The following semi-empirical formula (Equation~(\ref{eq:P0_tilda})) corrects this effect. Tested against numerical simulations, it gives a better estimate of $P_0$ to within $\sim10\%$ error in the range of CMF $\in 0.1\sim0.4$ and mass $\in 0.1 \sim 10 M_{\oplus}$. 

\begin{equation}
\boxed{\widetilde{P_{\rm 0}} \approx (200 + 600 \cdot {\rm CMF}) \cdot \left( \frac{M_{\rm p}}{M_{\oplus}} \right)~\rm GPa}
\label{eq:P0_tilda}
\end{equation}

\noindent A better approximation for the core pressure profile using $\widetilde{P_{\rm 0}}$ from Equation~(\ref{eq:P0_tilda}) and $P_{\rm CMB}$ from Equation~(\ref{eq:Pcmb}), shown as the purple curves in Figure~\ref{figure_profiles} Panel (3)a-d, is

\begin{equation}
\widetilde{P_{\rm 0}}(r) = \widetilde{P_{\rm 0}} - (\widetilde{P_{\rm 0}} - P_{\rm CMB}) \cdot \left( \frac{r}{R_{\text{core}}} \right)^2
\label{eq:Pcore_tilda}
\end{equation}

\noindent The corrected core gravity $\widetilde{g_{\rm core}(r)}$, shown as purple curves in Figure~\ref{figure_profiles} Panel (1)a-d, and the corrected core density $\widetilde{\rho_{\rm core}}$, shown as the purple curves in Figure~\ref{figure_profiles} Panel (2)a-d, can be calculated as from $\widetilde{P_{\rm 0}}$ as

\begin{equation}
\frac{\widetilde{g_{\rm core}(r)}}{g_{\rm core}(r)}=\frac{\widetilde{\rho_{\rm core}}}{\rho_{\rm core}}=\sqrt{\left( \frac{\widetilde{P_{\rm 0}}-P_{\rm CMB})}{P_{\rm 0}-P_{\rm CMB})} \right)}
\label{eq:gravitydensity_tilda}
\end{equation}

\noindent $\widetilde{\rho_{\rm core}}$ from Equation~(\ref{eq:gravitydensity_tilda}) tends to better approximates the core density near the center, while $\rho_{\rm core}$ from Equation~(\ref{eq:densitycore}) approximates the core density near the CMB. 

\section{Energy of Core Formation}

The energy of core formation can be estimated as the difference in gravitational energies between the uniform-density state and this simple analytical model.
According to the Virial Theorem~\citep{Haswell:2010}, the total gravitational energy is

\begin{equation}
E_{\text{grav}}=-3 \int_{\text{center}}^{\text{surface}}\frac{P}{\rho} \cdot dm
\label{eq:virial1}
\end{equation}

\noindent With the analytic forms of $P$ and $\rho$ in this model, Equation~(\ref{eq:virial1}) can be integrated to obtain

\begin{equation}
E_{\text{grav}}=-\frac{GM_{\rm{p}}^2}{3R_{\rm{p}}} \cdot \left(2-\frac{1}{5} \cdot \text{CRF}^3 \right)
\label{eq:Egrav}
\end{equation}

\noindent Comparing it to the gravitational energy of a uniform-density sphere, $E_{\textrm{grav, uniform sphere}}=-\frac{3}{5}\frac{GM_{\rm{p}}^2}{R_{\rm{p}}}$, the difference of the two can be regarded as the energy released during core formation (gravitational energy released from the concentration of denser materials toward the center): 

\begin{equation}
E_{\textrm{diff}}=E_{\textrm{grav, uniform sphere}}-E_{\text{grav}}=\frac{GM_{\rm{p}}^2}{R_{\rm{p}}} \left(\frac{1}{15} (1-\text{CRF}^3) \right) = \frac{GM_{\rm{p}}^2}{R_{\rm{p}}} \left(\frac{1}{15} (1-\text{CMF}^{3/2}) \right)
\label{eq:Ediff1}
\end{equation}

\noindent Since CMF $\in 0.1\sim0.4$, the term $\text{CMF}^{3/2}$ is small enough to be dropped to give

\begin{equation}
\boxed{E_{\textrm{diff}} \approx \frac{1}{15}\frac{GM_{\rm{p}}^2}{R_{\rm{p}}} \approx \frac{1}{10} \mid E_{\text{grav}}\mid}
\label{eq:Ediff2}
\end{equation}

Therefore, the energy released during core formation is $\sim10\%$ of the total gravitational energy of such a rocky planet. For Earth, $E_{\textrm{diff}, \oplus} \approx 2.5*10^{31}$ J. The calculated values of other planets are listed in Table~\ref{Table1}.

\section{Thermal Content of the Planet}

Since the temperatures inside the mantle of such a planet are likely above the Debye temperature of the solid, the heat capacity per mole of the atoms can be approximated as $3R$ (gas constant $R=8.314~\rm J~K^{-1} mol^{-1}$). The specific heat capacity (heat capacity per unit mass) is $3R/\mu$ where $\mu$ is the average atomic weight of the composition, which for Mg-silicates (MgO, SiO$_2$, or any proportion of them combined, such as MgSiO$_3$ or Mg$_2$SiO$_4$) is 0.02$\rm~kg~mol^{-1}$. The specific thermal energy $u_{\text{th}}$ of the mantle material is thus

\begin{equation}
u_{\text{th}}=\frac{3RT}{\mu}
\end{equation}

\noindent where $T$ is temperature. The total thermal energy of mantle is calculated by the integration: 

\begin{equation}
E_{\text{th,mantle}}=\int^{M_{\rm{p}}}_{M_{\text{core}}} u_{\text{th}} \cdot dm = M_{\rm{p}} \cdot \int^{1}_{\text{CMF}} u_{\text{th}} \cdot dx
\end{equation}

\noindent where $x \equiv \frac{m}{M_{\rm{p}}}$. In this model, the mantle density $\rho_{\text{mantle}} \propto \frac{1}{r} \propto \frac{1}{\sqrt{m}}$ (Equation~(\ref{eq:densitymantle})). On the other hand, with the assumption of the adiabatic temperature gradient in the mantle and the introduction of the Gr\"{u}neisen parameter $\gamma \equiv \frac{\partial ln(T)}{\partial ln(\rho)} |_{adiabat}$, the specific thermal energy can be rewritten to show its functional dependence on density $\rho$ or mass $m$~\citep{Zeng:2016b}:

\begin{equation}
u_{\text{th}}=\frac{3R \cdot T_{\text{mp}}}{\mu} \cdot \left(\frac{\rho}{\rho_0}\right)^{\gamma} = \frac{3R \cdot T_{\text{mp}}}{\mu} \cdot \left(\frac{m}{M_{\rm{p}}}\right)^{-\frac{\gamma}{2}}
\end{equation}

\noindent where $T_{\text{mp}}$ (mantle potential temperature) is defined as the temperature where the mantle adiabat is extrapolated to zero pressure. For silicates, $\gamma \sim 1$, then,  

\begin{equation}
E_{\text{th,mantle}} = 2 \cdot M_{\rm{p}} \cdot \frac{3R \cdot T_{\text{mp}}}{\mu} (1-\sqrt{\text{CMF}})
\label{eq:mantleEth}
\end{equation}

\noindent For CMF $\approx 0.3$, $\sqrt{\text{CMF}} \approx \text{CRF} \approx 0.5$, and the total thermal energy of the mantle (considering only vibrations of the atoms in crystal lattices, while neglecting the electron contribution) is

\begin{equation}
E_{\text{th,mantle}} \approx \frac{3R \cdot T_{\text{mp}}}{\mu} \cdot M_{\rm{p}} \approx \left(\frac{T_\text{mp}}{1000\text{K}} \right) \cdot \left( \frac{M_{\rm{p}}}{M_{\oplus}} \right) \cdot 7 \cdot 10^{30} \rm~J
\label{eq:mantleEth2}
\end{equation}

\noindent Equation~(\ref{eq:mantleEth2}) suggests that concerning the thermal content, the mantle can be treated as an uncompressed mass of $M_{\rm p}$ at isothermal temperature of $T_{\text{mp}}$. Therefore, an effective heat capacity $C_{\text{th,mantle}}$ of the mantle can be defined with respect to $T_\text{mp}$:

\begin{equation}
C_{\text{th,mantle}} \approx \frac{3R}{\mu} \cdot M_{\rm{p}} \approx \left( \frac{M_{\rm{p}}}{M_{\oplus}} \right) \cdot 7 \cdot 10^{27} \rm~J~K^{-1}
\label{eq:heatcapacity}
\end{equation}

\noindent The detailed calculation in~\citet{Stacey:2008} shows that the effective heat capacity of the Earth's mantle is $7.4 \cdot 10^{27}\rm~J~K^{-1}$, indeed close to our estimate. Since the core is smaller in comparison (CMF $\in 0.1\sim0.4$) and the core material has smaller specific heat capacity than silicates, the thermal content of the core should be generally less than that of the mantle. The mantle shall dictate the cooling of the core~\citep{Stacey:2008}. Therefore, the mantle heat capacity can be regarded as an approximation for the total heat capacity of a planet. 

Due to a feedback mechanism of silicate melting, there is good reason to set $T_{\text{mp}} \approx 1600 \rm~K$ for Earth and super-Earths for a first approximation~\citep{Stixrude:2014}. Then the temperature profiles inside these planets can be estimated using formulae in~\citet{Stixrude:2014}. The results are listed in Table~\ref{Table1} and plotted in Figure~\ref{figure_profiles} Panel (4)a-d. In general, the mantle adiabat has shallower slope than the mantle melting curves~\citep{Zeng:2016b}, so the mantle is mostly solid, while the core is partially or fully molten. 

\section{Moment of Inertia}

The moment of inertia is calculated from Equation~(\ref{eq:MoI1}), where $x$ represents the distance of the mass element to the rotational axis and the integration is over the entire volume ($V$): 

\begin{equation}
I = \iiint\limits_V x^2 \cdot \rho(\vec{r}) \cdot  dV
\label{eq:MoI1}
\end{equation}

\noindent Considering two simple cases: 

$\begin{cases}
\text{thin spherical shell with radius $R_{\rm{p}}$, } I_{\text{shell}} = \frac{2}{3}MR_{\rm{p}}^2\\
\text{uniform solid sphere with radius $R_{\rm{p}}$, } I_{\text{solid sphere}} = \frac{2}{5}MR_{\rm{p}}^2
\end{cases}$

Defining $C\equiv I/MR_{\rm{p}}^2$ as the moment-of-inertia factor, for shell $C=\frac{2}{3}$, and for sphere $C=\frac{2}{5}$. Smaller $C$ corresponds to the mass being more concentrated toward the center. For this model, the moment of inertia of the core and the mantle can each be calculated separately, then combined to give the total moment of inertia of the planet: 

\begin{equation}
I_{\textrm{core}} = \frac{2}{5}M_{\rm{p}}R_{\rm{p}}^2\cdot \text{CMF}^2
\label{eq:MoIcore}
\end{equation}

\begin{equation}
I_{\textrm{mantle}} = \frac{1}{3}M_{\rm{p}}R_{\rm{p}}^2\cdot (1-\text{CMF}^2)
\label{eq:MoImantle}
\end{equation}

\begin{equation}
I_{\text{total}} = I_{\textrm{core}} + I_{\textrm{mantle}} = \frac{1}{3}M_{\rm{p}}R_{\rm{p}}^2\cdot \left( 1+\frac{1}{5} \cdot \text{CMF}^2 \right)
\label{eq:MoIcombined}
\end{equation}

Considering CMF$\in 0.1\sim 0.4$,  the term $\frac{1}{5}\text{CMF}^2$ can be ignored, so $C \approx \frac{1}{3}$. In the solar system, the $C$ for Mercury, Venus, and Earth is indeed very close to $\frac{1}{3}$~\citep{Rubie:2007}. Here we show that $C \approx \frac{1}{3}$ can be generalized to other rocky exoplanets:  

\begin{equation}
\boxed{I_{\text{planet}} \approx \frac{1}{3} \cdot M_{\rm{p}} \cdot R_{\rm{p}}^2}
\label{eq:moment_of_inertia}
\end{equation}

For Earth, $M_{\oplus} R_{\oplus}^2 = 2.4 \cdot 10^{38}\rm~kg~m^2$, and $I_{\oplus} \approx \frac{1}{3} M_{\oplus} R_{\oplus}^2 = 8 \cdot 10^{37}\rm~kg~m^2$. The angular momentum of Earth's rotation is $L_{\oplus} = I_{\oplus} \cdot \Omega_{\oplus} = 6 \cdot 10^{33}\rm~kg ~m^2~s^{-1}$. The total rotational energy of Earth is $E_{\text{rot}} = \frac{1}{2} \cdot I_{\oplus} \cdot \Omega_{\oplus}^2 = \frac{L_{\oplus}^2}{2 \cdot \Omega_{\oplus}^2} =2 \cdot 10^{29}$ J, where $\Omega_{\oplus} = 7.3 \cdot 10^{-5}\rm~rad~s^{-1}$ is the angular frequency of Earth's rotation. The calculated moment of inertia of other planets are shown in Table~\ref{Table1}.

\section{Conclusion}

A simple analytical model for rocky planetary interiors is presented here and compared to numerical results. 
It explores the scaling relations among the following five aspects: (1) the relative size and mass of the core and the mantle, (2) the interior pressure and gravity, (3) the core formation energy and the gravitational energy, (4) the heat content and temperature profiles, and (5) the moment of inertia. Other results can be derived from this model. 

Although being approximate, these results are straightforward to apply, as in many cases mass and radius are only measured approximately. Combined with the mass-radius relation, these formulae shall provide us with a new way of looking at the rocky planetary interiors, complementing the numerical approach.

\section{Acknowledgement}
This work was supported by a grant from the Simons Foundation (SCOL [award \#337090] to L.Z.). The authors would like to thank Mr. and Mrs. Simons and the Simons Foundation for generously supporting this research project. The author L.Z. would like to thank his father Lingwei Zeng and mother Ling Li for continuous support and help. He also would like to thank Eugenia Hyung for insightful discussion of this paper. 
Part of this research was also conducted under the Sandia Z Fundamental Science Program and supported by the Department of Energy National Nuclear Security Administration under Award Numbers DE-NA0001804 and DE-NA0002937 to S.B.J (PI) with Harvard University. 
This research reflects the authors' views and not those of the DOE. Finally, the authors would like to thank Dimitar D. Sasselov for insightful suggestions and helpful comments on this paper. 

\clearpage


\bibliographystyle{apj}
\bibliography{mybib}

\end{CJK*}
\end{document}